# Formation of ripple pattern on silicon surface by grazing incidence ion beam sputtering

S. A. Mollick and D. Ghose

*Saha Institute of Nuclear Physics, Sector – 1, Block – AF, Bidhan Nagar, Kolkata 700 064, India*

## Abstract

Off-normal low energy ion beam sputtering of solid surfaces often leads to morphological instabilities resulting in the spontaneous formation of ripple structures in nanometer length scales. In the case of Si surfaces at ambient temperature, ripple formation is found to take place normally at lower incident angles with the wave vector parallel to the ion beam direction. The absence of ripple pattern on Si surface at larger angles is due to the dominance of ion beam polishing effect. We have shown that a gentle chemical roughening of the starting surface morphology can initiate ripple pattern under grazing incidence ion beam sputtering, where the ripple wave vector is perpendicular to the ion beam direction. The characteristics of the perpendicular mode ripples are studied as a function of pristine surface roughness and ion fluence. The quality of the morphological structure is assessed from the analysis of ion induced topological defects.

# I. INTRODUCTION

The present day electronic devices are fabricated mostly on Si platforms which need patterning and texturing of surfaces at nanometer length scales [1, 2]. Recently, it has been shown that ion-beam-sputtering (IBS) induced Si rippled surface can be useful as a growth template for the production of metal nanodot arrays or wire-like patterns with long range order and tunable periodicity [3], which have important implications in some technological areas, e.g. in the fabrication of data storage systems [4, 5] and optoelectronic devices [6]. According to the Bradley and Harper (BH) theory of pattern formation [7], the periodic wavelike structure developed by IBS is the result of the interplay between the curvature-dependent sputtering and various surface relaxation processes. BH also predicted that the pattern will be oriented with its wave vector either in the x-direction (parallel mode ripples) or y-direction (perpendicular mode ripples) depending on whether the incident beam angle is smaller or larger than a critical angle. Here, the beam direction is supposed to be in the x-direction. However, current literature [8, 9] shows only the formation of x-direction ripples in single crystal Si wafers at room temperature. This is quite surprising since the y-direction erosion-induced instabilities are shown to be very robust at all beam incident angles for a wide class of erosive response functions including Gaussian and non-Gaussian [10]. The other characteristics of the y-direction ripples are that they are stationary waves during the IBS growth process as the gradient-dependent erosion operates only in the transverse direction [7] and secondly, the ripples can be well elongated along the beam direction, e.g. see ref. [11]. Such a periodic pattern may be helpful to improve the alignment of the deposited nanometal structures at the thermodynamically stable positions, e.g. valleys of the templated substrates [3, 12]. Recently, Zhang et al. [13] reported the formation of perpendicular mode ripples on Si (100) wafers under 5 keV $Xe^+$ bombardment at high angles of ion incidence ($\geq 80^0$);

unfortunately, the waviness of the developed topography was difficult to recognize. The reason for the absence of clear ripple topography in their experiment, might be the use of a smooth Si surface (rms roughness = 0.05 nm), where the minimum pre-surface roughness required for ripple formation can probably not be achieved by the sputter noise present in the beam because of the dominance of ion beam smoothing effect at grazing incidence sputtering [14]. Very recently, we have shown that gentle roughening of the as-received flat Si surface by chemical etching can shorten the ion beam fluence necessary for parallel mode ripple formation and also such a pre-roughened surface acts as a precursor to the formation of regular nanostructures [15]. Our earlier works on metal films [16, 17] also showed evidences about the role of topographic roughness in determining pattern formation. The present article demonstrates that it is possible to create perpendicular mode ripples on single crystal Si wafers subjected to low energy grazing incidence ion beam sputtering at ambient temperature, provided that the initial surface is gently roughened by chemical etching. Specifically, the morphology of the perpendicular mode ripples is investigated by varying the starting surface roughness and ion fluence in order to explore the regime where the optimization of the pattern morphology can be achieved. We have also studied the degree of ordering of the ripple structure by quantifying the structural defects from the topographic images.

**II. EXPERIMENTAL**

Commercially available mirror-polished Si (100) wafers, after appropriate degreasing and cleaning, was chemically etched in a solution of 5mol/L $HNO_3$ and 12mol/L HF at room temperature in order to prepare rough surface morphology of different degree by varying the etching time. A number of etched samples with different *rms* roughness, w were prepared. The rms roughness of the as received Si (100) samples was ~ 0.1 nm. The IBS of the etched samples was carried out in a low energy ion beam set-up [18] under 16.7 keV $O_2^+$ bombardment at different incident angles, θ ranging from $30^0$ to $85^0$ with respect to the surface normal. The beam current density was around 10μA/cm$^2$. The topography of the samples was examined by a MMSPM NanoScope IV (Veeco, USA) in contact mode at ambient conditions. Fig. 1(c) shows a typical virgin topography of the chemically etched samples which is characterized by asperities of different sizes possibly due to non-uniform etching. The wavelength of the ripples developed by IBS on Si was determined either from autocorrelation function or from fast Fourier transformations (FFTs) using the atomic force microscopic (AFM) images.

## III. RESULTS AND DISCUSSION

Fig. 1 shows some typical AFM images of the sputtered topography on the flat Si and the etched Si samples as a function of ion incident angles. For the flat surface [Figs. 1(a) - (b)], only parallel mode ripples are found to develop in a narrow angle window close to $45^0$ - $60^0$ in agreement with our earlier results [19]. The situation is dramatically changed for the pre-roughened surface, where both parallel and perpendicular mode ripples develop under the same bombardment conditions [Figs. 1(c) – (f)]. In addition, one can find the threshold angle ($θ_c = 65^0$), above which the ripple pattern rotates by $90^0$. At this angle, blunt hill-like surface structures are found to form indicating the suppression of anisotropic sputtering [cf., Fig. 1(e)]. The FFTs shown in the inset of the respective AFM images are in conformity with the ordering and the orientations of the pattern with respect to the beam direction at different incident angles for the as-received as well as for the etched Si surfaces.

The evolution of perpendicular mode ripples, which is the present study of interest, for the case of etched Si, with fluence $\phi$ ranging from $5 \times 10^{16}$ to $1.5 \times 10^{18}$ O atoms/cm$^2$ is shown in Fig. 2. At low fluence [Fig. 2(b), $\phi = 5 \times 10^{16}$ O atoms/cm$^2$], the surface exhibits tiny cone-like structures inclined in the beam direction; at increasing fluence, these structures overlap leading to a wavy pattern. The functional dependence of the lateral periodicity (wavelength) and the vertical height fluctuations (amplitude) of the developed morphology on the fluence are summarized in Figs. 3 (a) and (b). Initially the wavelength decreases with fluence followed by a minimum and then increases again. On the other hand, the amplitude of the pattern reduces with fluence consistent with the expectations from ion beam polishing effect at grazing incidence sputtering. In another experiment, the development of ripple structure as a function of pristine surface roughness was investigated for the incident angle held at $80^0$ and the fluence fixed at $4 \times 10^{17}$ O atoms/cm$^2$. The results show an increase of the ripple wavelength with the roughness of the samples [Fig. 3 (c)].

The BH model assumed a perturbed planar starting surface characterized by the superposition of periodic height modulations with a broad range of spatial frequencies [7]. The individual Fourier components of the surface height are amplified with sputter time according to the BH dispersion relation and finally a characteristic spatial frequency is dominated for which the growth rate is maximized. In the later refinements of the BH model, the noisy characteristic of the incident beam is included, which is identified as the driving force for generating random small surface undulations on the initial flat surface during the early stages of bombardment [20, 21]. Although the presence of sputter noise can readily explain the formation of parallel mode ripples on an initially smooth Si surface at low oblique beam angles [cf., Fig. 1(b)], the absence of perpendicular mode ripples on the same Si surface under grazing ion incidence indicates that the threshold roughness to trigger the ripple formation may not always be achieved by the beam noise only. In fact, Kimura et al. [22] showed that the glancing angle keV ion beam sputtering ($\theta > 75^0$) actually reduces the rms surface roughness of a Si (001) wafer (with initial rms roughness $\leq 0.4$ nm) to the order of a single atomic step height, i.e. the incident ion beam effectively polishes the surface to a high degree of smoothness. In other words, for the as-received Si surface, the resultant surface roughness that developed due to competition between stochastic roughening and ion beam smoothing seems to be far below the requirement for the ripple nucleation.

The BH theory [7] basically includes the first order (angle or gradient dependent sputtering) and higher order processes (curvature dependent sputtering and smoothing) for the topography evolution. At near glancing sputtering, the BH expression for the time evolution of the surface height, $h(x,y)$, can be written as

$$\frac{\partial h}{\partial t} = -\frac{fY_0(\theta)}{n}\cos\theta + K\frac{\partial^2 h}{\partial y^2} - B\nabla^2(\nabla^2 h), \qquad (1)$$

where $f$ is the incident projectile flux, $Y_0(\theta)$ is the sputtering yield at an angle $\theta$ to the surface normal and $n$ is the number density of target atoms. The factor $K$ describes the sputter yield dependence on the surface curvature that leads to roughening and $B$ describes the role of smoothening by surface diffusion. Additionally one should take into considerations the shadowing of ion trajectories by surface structures and reflection of incoming ions, which are increasingly important at larger angles of incidence.

The typical AFM image of the etched Si surface and its FFT spectrum [e.g., Fig. 1 (c)] show the existence of both fine scale and large scale surface structures at different frequency ranges. For such cases, both macroscopic (first order) and microscopic (higher order) erosion processes are simultaneously activated on the surface. At the beginning of ion bombardment, the small surface irregularities with higher spatial frequencies are smoothed out rapidly via different smoothing or relaxation mechanisms [14], which include curvature-dependent surface diffusion, viscous flow in the damaged near-surface region and the so-called ion-induced surface diffusion arising from higher order (fourth order) spatial derivatives of local surface height [20]. On the other hand, the longer wavelength taller asperities are evolved mainly by the first order erosion processes [the first term in Eq. (1)]. Since $Y_0(\theta)$ is a function of $\theta$, there will be a change of surface slope with sputter time and it is most likely that the macro-asperities are evolved to cone-like structures inclined along the ion beam direction [23]. Formation of cones or pyramids on ion-bombarded surfaces is a well-known phenomenon and several alternative processes to explain their formation can be found in the literature [24]. The cones formed this way are unstable and decay faster with beam exposure time and some kind of leveling of the surface at the intermediate stage of sputtering can be achieved. It is noteworthy that such angle-dependent ion beam smoothening is less efficient compared to those involved for smoothing of fine scale surface structures [25]. At this stage, the peak-to-valley distance of the surface features is decreased enough so that the first order erosion instability loses its importance and there is a crossover to curvature-dependent erosion leading to the development of the fine scale BH-ripple morphology. Toma et al. [26, 27] suggested that, the initial pattern wavelength [cf., Fig. 3(a)], which is generally found wider, is governed by the spatial separation of the larger asperities via the process of ion beam shadowing. With increasing fluence, this asperity-size dependent wavelength decreases until a threshold fluence is reached when the BH mechanism for ripple formation takes over. The coarsening of BH-ripple wavelength due to further increase of fluence can be addressed following a hydrodynamical approach [28] which considers the temporal evolution of the surface height coupled with the growth of redeposited sputtered atom layer. Toma et al. [26, 27] also showed that the transition fluence for the onset of BH-ripples is strongly correlated to the substrate roughness; this actually shifts to higher values for higher initial roughness and asperity size. This is probably the reason for obtaining larger pattern wavelength at higher pristine roughness for a fixed bombarding fluence [Fig. 3(c)]. In this context, we would like to point out that the present experiment as well as our previous experiments with metal films at grazing ion bombardment condition [16, 17] reveals a decrease of roughness with sputter time followed by saturation at lower level than the initial roughness [cf., Fig. 3(b)]. The results are not unlikely because the vertical growth of the topography is limited by the efficient ion beam shaving processes operating at high angles of incidence [14, 29].

Another interesting aspect of ripple structures is the development of defects in the form of broken, merged or bended lines, which were also revealed in the computer simulated wavy patterns as studied in Refs. [30, 31]. Such defects are thought to associate with the randomness of the initial surface morphology and also can arise due to the stochastic nature of the formation process. One can quantitatively estimate the topological defects from the AFM images following the procedures described in Refs. [31, 32]. Briefly, a flattened AFM topograph is first reduced to a binary image using a suitable threshold algorithm and then thinned to line traces of single pixel width. The defects are counted where a single line has a branching or truncation at its ends. In order to decouple from the wavelength change, the defect number is normalized by multiplying with the square of wavelength and dividing with the image area. In passing, we mention that the previous authors [31, 32] have used the Otsu's algorithm [33] for the height threshold determination, which is based on a global image thresholding appropriate to the multimodal type of images. For unimodal type of images like the present one, a different threshold algorithm [34], similar to that of Carson et al. [35], is developed that takes into account the local threshold for every pixel of images. The latter method generates the expected binary image without losing its structural information. A few representative thresholded binary images and their thinned versions corresponding to the AFM ripple topographies on different etched samples are shown in Fig. 4.

Figs. 5 (a) and (b) show the dependence of $N_D$ as a function of initial roughness and projectile fluence. $N_D$ generally increases with roughness for a particular fluence. On the other hand, for a given pristine roughness, $N_D$ is found to decrease sharply with fluence and the ripple topography evolves towards more ordered structure until $N_D$ reaches to a minimum level around $4 \times 10^{17}$ O atoms/cm$^2$; after that $N_D$ tends to increase again. Recently, Keller et al. [31] studied the dynamics of ripple defects by simulating ripple pattern on the Si surface. They showed that the pattern quality is strongly dependent on the bombarding fluence. The short ripples developed at the early time get longer with increasing time and coalesce with neighboring ripples forming defects in the junction either in the form of interstitials or bifurcations. Subsequently, these defects are self-annealed through pinching off processes, thereby reducing the defect density with increasing fluence. In the regions of higher defect density, however, the defects may segregate and form clusters, which may retard the fluence dependent annealing processes and can be dominant after certain fluence. It is found that samples of 3 - 7 nm rms roughness when bombarded with fluence $4 \times 10^{17}$ O atoms/cm$^2$ produces good quality perpendicular mode ripples with less topological defects.

## V. CONCLUSIONS

In summary, our findings suggest that a minimum degree of roughness ($\approx 2$ nm) of the starting Si surface morphology is required to trigger the curvature-dependent erosion instability at grazing incident angles (perpendicular mode BH ripples), which may not be achieved by the presence of stochastic noise of the incident beam only. The pre-roughened surface is also found to be beneficial to determine experimentally the transition angle for the ripple rotation. The topological defects developed in the perpendicular mode ripples were analyzed with an improved image threshold algorithm [34] that considers the multiscale features for thresholding. Under the present experimental conditions, the defect density is found to be minimal around the fluence $4 \times 10^{17}$ O atoms/cm$^2$ for the initial surface roughness in the range 3 – 5 nm. Such periodic and elongated furrows and ridges on the Si surface can be produced by ion beam sputtering over areas of several square millimeters in a fast and cost effective way and may be ideal as nanoscale growth templates for the physical synthesis of functional materials [9].

## ACKNOWLEDGEMENTS

We would like to thank Prof. Sandip Sarkar for his help in the defect analysis from AFM images. The involvements of Drs. P. Mishra and P. Karmakar in the experiment are gratefully acknowledged.

**Figure Captions:**

Fig. 1. AFM images ($10 \times 10$ μm$^2$) showing the morphology of a flat Si(100) surface (a) and the surface ripples on the flat Si surface sputtered at $60^0$ (b). AFM images showing the morphology of the chemically roughened Si (100) surface (c) and the morphology of the patterns formed on the roughened Si surface during sputtering for incident angles of $60^0$ (d), $65^0$ (e) and $80^0$ (f). The insets in the AFM images show the corresponding FFTs. The arrow indicates the beam direction.

Fig. 2. (a) - (f) Evolution of topographic pattern with increasing fluence. The sample [corresponding to Fig. 1(c)] was sputtered at θ = $80^0$ by 16.7 keV $O_2^+$ ions. The scan area of each image is $10 \times 10$ μm$^2$. The arrow indicates the beam direction.

Fig. 3. (a) and (b) show, respectively, the plots of ripple wavelength and rms roughness versus projectile fluence and (c), the plot of the ripple wavelength as a function of pristine surface roughness of the samples. The sample was sputtered at θ = $80^0$ by 16.7 keV $O_2^+$ ions.

Fig. 4. $10 \times 10$ μm$^2$ etched Si samples bombarded at θ = $80^0$ by 16.7 keV $O_2^+$ ions. (a) AFM topography of the rippled surface; the topography of the pristine surface is shown in the inset, (b) thresholded image, and (c) thinned version.

Fig. 5. (a) and (b) are, respectively, the plots of normalized defect density for ripple morphology versus pristine roughness and projectile fluence. The samples were sputtered at θ = $80^0$ by 16.7 keV $O_2^+$ ions.

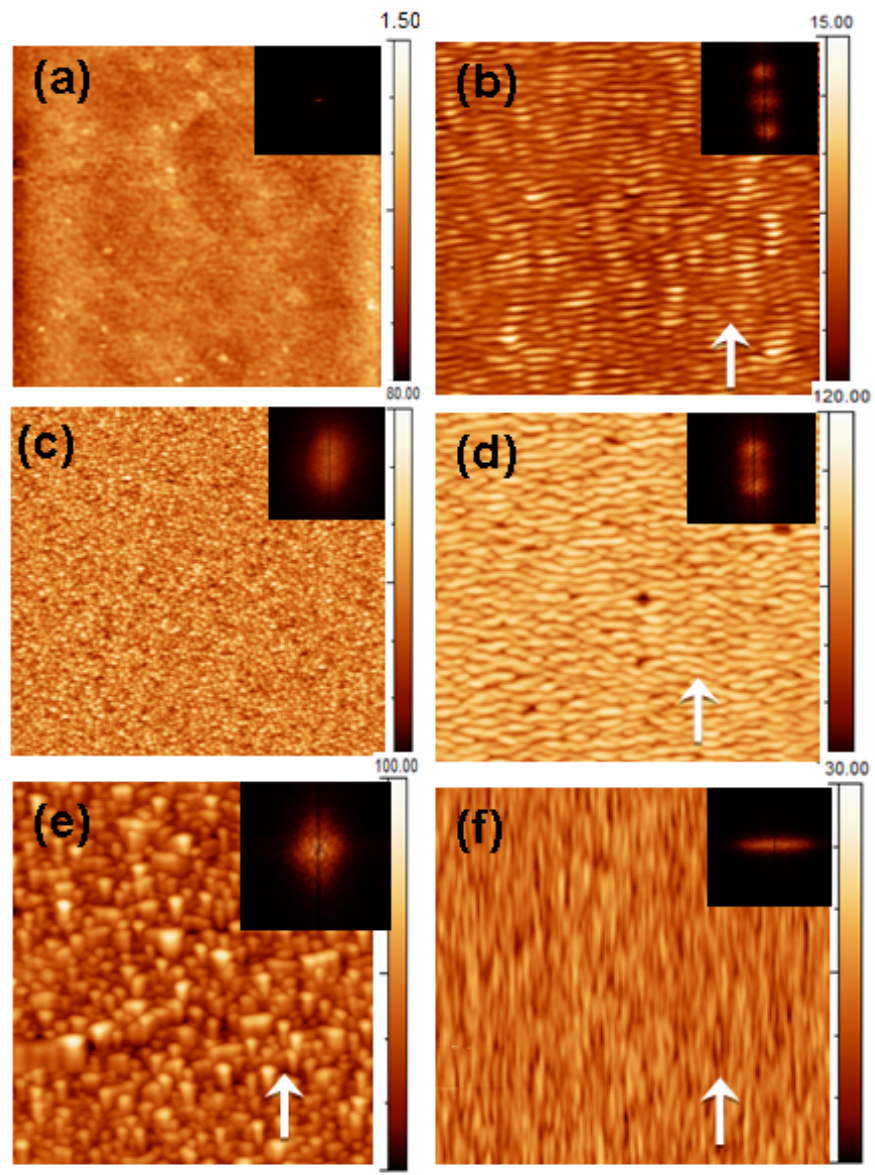

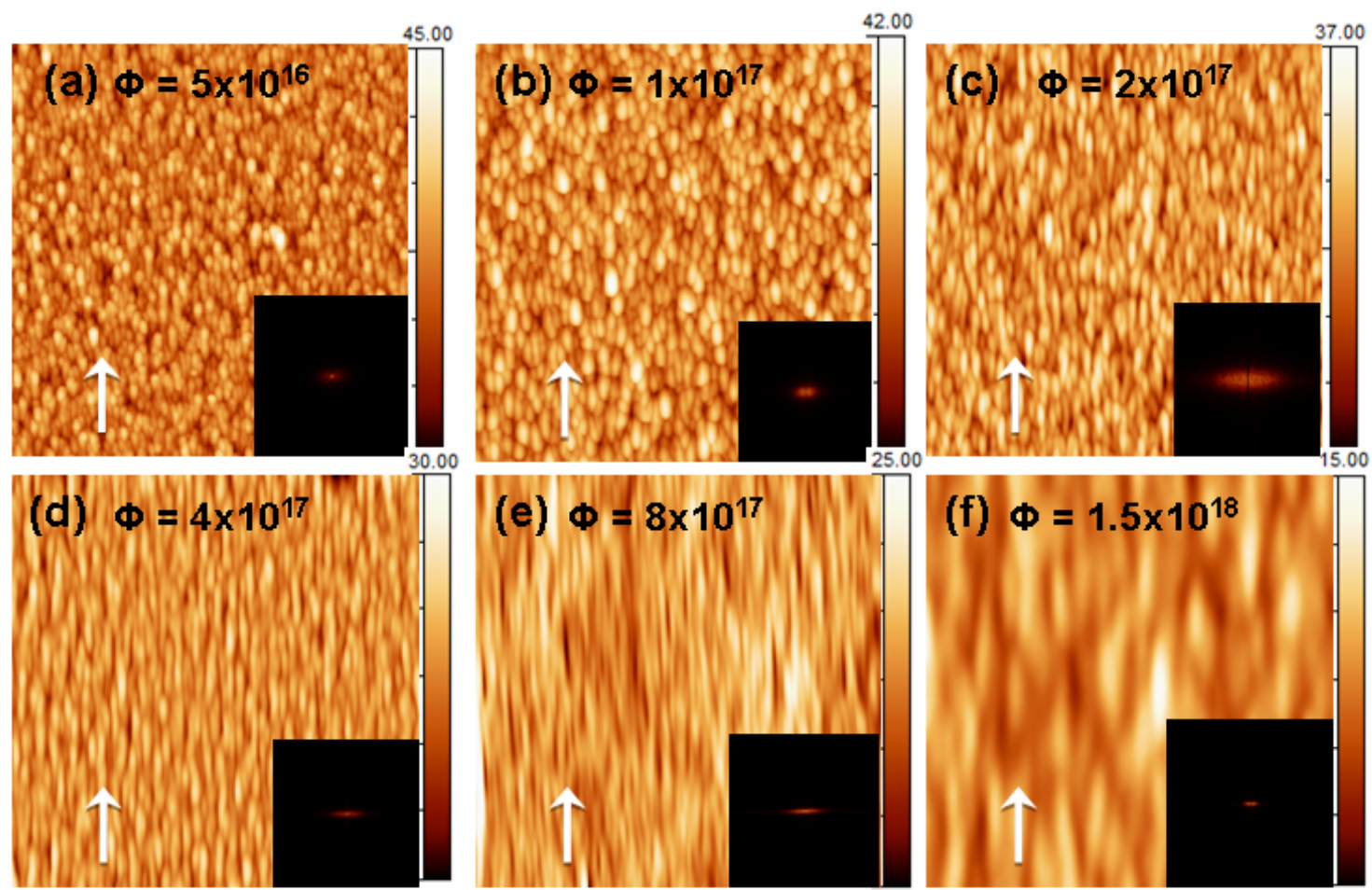

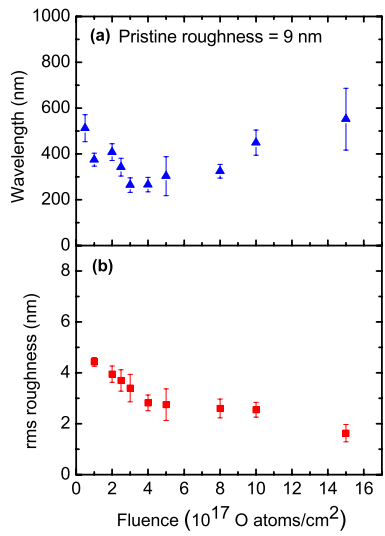
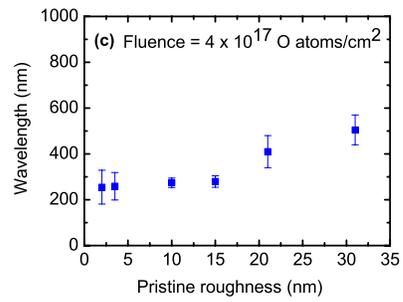

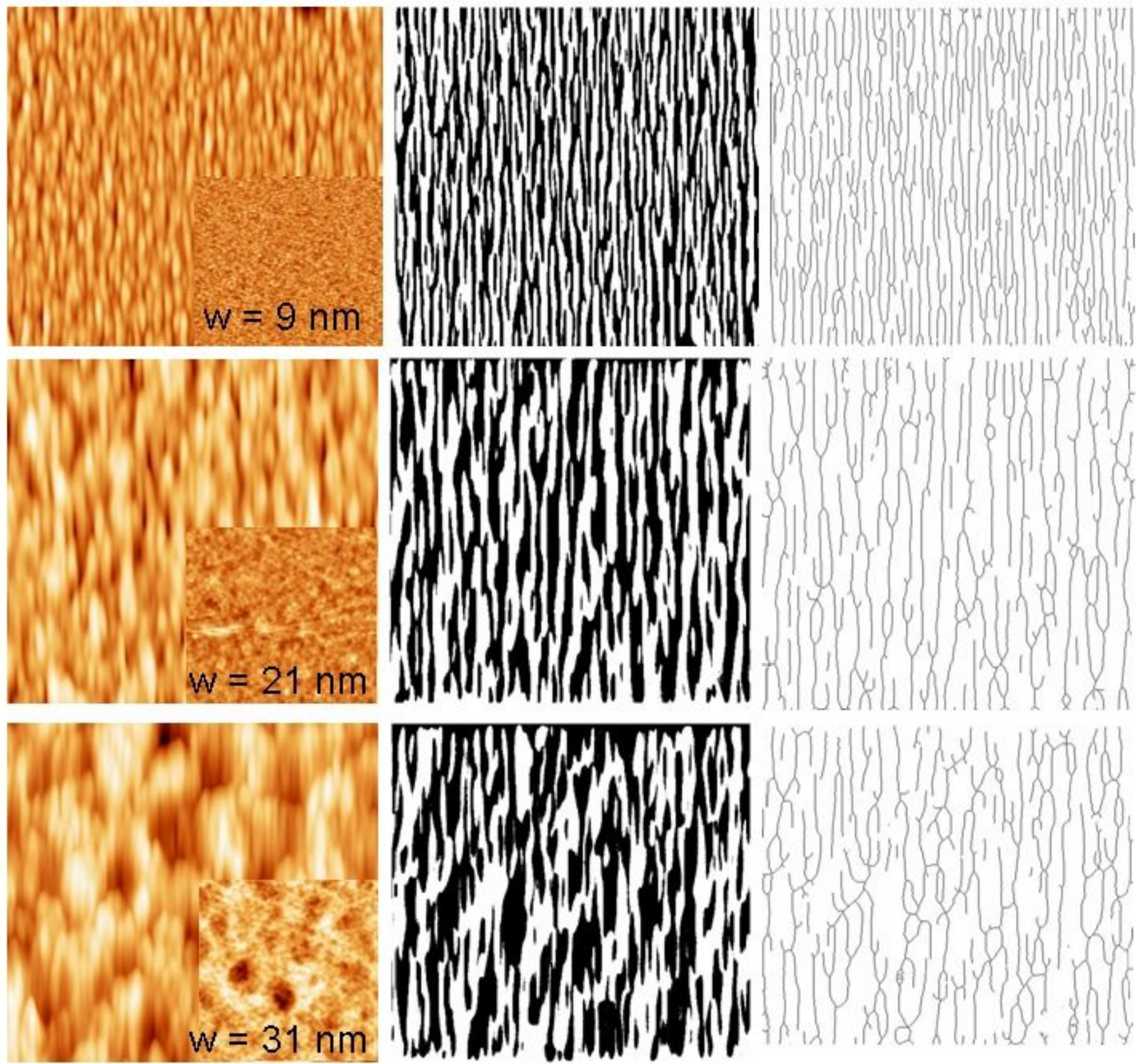

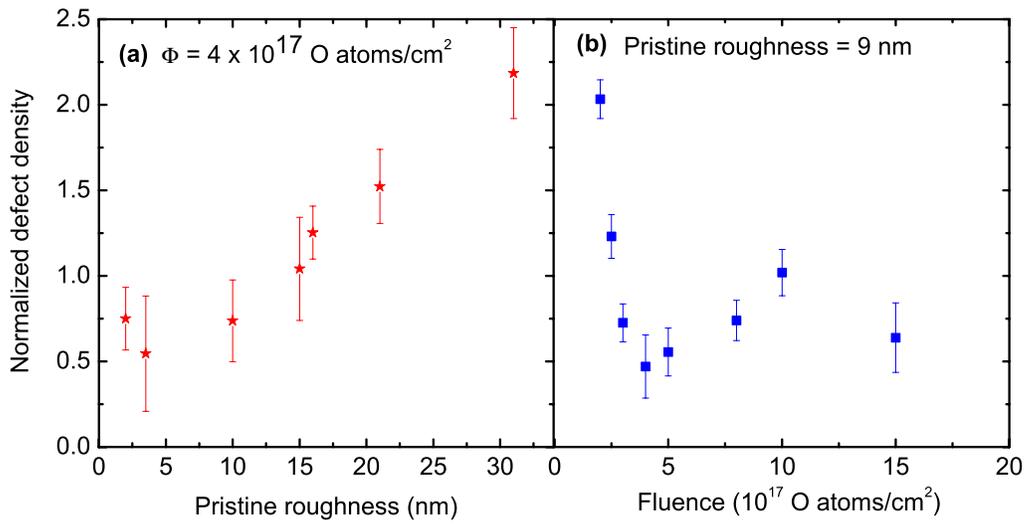